\documentstyle[referee]{laa}

\begin{document}
\thesaurus{07  
	   (07.09.1; 
	    07.13.1;  
	   )}

\title{Aberration of light and Motion of Real Particle}
\author{J.~Kla\v{c}ka}
\institute{Institute of Astronomy,
   Faculty for Mathematics and Physics, Comenius University \\
   Mlynsk\'{a} dolina, 842~48 Bratislava, Slovak Republic}
\date{}
\maketitle

\begin{abstract}
Correct and complete (to terms of $\vec{v} / c$ -- $\vec{v}$ is
particle's velocity, $c$ is the speed of light) derivation of
equation of motion for real dust particle under the action of
electromagnetic radiation is derived. The effect of aberration of light is
used. Equation of motion is expressed in terms of
particle's optical properties, standardly used in optics for stationary
particles.

\keywords{cosmic dust, interplanetary dust, aberration of light}

\end{abstract}

\section{Introduction}
Equation of motion for perfectly absorbing spherical dust particle
under the action of electromagnetic radiation was derived by
Robertson (1937). Relativistic generalization for the case when scattered
radiation is in the direction of the incident radiation -- in proper reference
frame of the particle -- was presented by Kla\v{c}ka (1992, 2000a).

However, real particles scatter radiation in a more complicated manner.
The consequence of this reality may be a completely different orbital
evolution of dust particles. Kocifaj and Kla\v{c}ka (1999) show this
for really shaped stationary rapidly rotating particle.
Relativistically covariant equation of motion for real cosmic dust
particle was presented in Kla\v{c}ka (2000b).

We want to derive equation of motion for real dust particle
under the action of electromagnetic radiation. We want to make the
explanation as simple as possible -- to be understandable to majority
of scientific community working in the field of interplanetary matter.
The effect of aberration of light will be used, the ideas are
concentrated in  Kla\v{c}ka (1993a, 1993b).

\section{Proper reference frame of the particle -- stationary particle}
The term ``stationary particle'' will denote particle which does
not move in a given inertial frame of reference,
although we admit its rotational
motion around an axis of rotation (with negligible rotational velocity).
Primed quantities will denote quantities measured in the
proper reference frame of the particle.

The flux density of the radiation energy
of photons scattered into an elementary solid angle
$d \Omega ' = \sin \vartheta ' ~ d \vartheta ' ~ d \varphi '$
is proportional to  $p' ( \vartheta ', \varphi ') ~ d \Omega '$, where
$p' ( \vartheta ', \varphi ')$ is ``phase function''.
Phase function depends
on orientation of the particle with respect to the direction of the
incident radiation and on the particle characteristics;
angles $\vartheta '$, $\varphi '$ correspond to the direction (and orientation)
of travel of the scattered radiation, $\vartheta '$ is polar angle
and it equals zero for the case of the travel of the ray in the orientation
identical with the unit vector $\hat{\vec{S}_{i}} '$ of the incident radiation.
The phase function fulfills the condition
\begin{equation}\label{1}
\int_{4 \pi} p' ( \vartheta ', \varphi ')~ d \Omega ' = 1 ~.
\end{equation}

The momentum of the incident beam of photons which is lost in the process
of its interaction with the particle is proportional to the cross-section
$C'_{ext}$ (extinction). The part proportional to $C'_{abs}$ (absorption)
is completely lost and the part proportional to
$C'_{ext} ~-~ C'_{abs} = C'_{sca}$ (scattering) is again reemitted.

The momentum (per unit time) of the scattered photons into an elementary
solid angle $d \Omega '$ is
\begin{equation}\label{2}
d \vec{p'}_{sca} = \frac{1}{c} ~ S' ~ C'_{sca} ~
		   p' ( \vartheta ', \varphi ')~ \hat{\vec{K'}} ~ d \Omega ' ~,
\end{equation}
where
\begin{equation}\label{3}
\hat{\vec{K'}} = \cos \vartheta '~\hat{\vec{S}_{i}} ' ~+~
		 \sin \vartheta ' ~ \cos \varphi ' ~ \hat{\vec{e}_{1}} ' ~+~
		 \sin \vartheta ' ~ \sin \varphi ' ~ \hat{\vec{e}_{2}} ' ~.
\end{equation}
$S'$ is the flux density of radiation energy. The system of unit vectors
used on the RHS of the last equation forms an orthogonal base.
The total momentum (per unit time) of the scattered photons is
\begin{equation}\label{4}
\vec{p'}_{sca} = \frac{1}{c} ~ S' ~ C'_{sca} ~ \int_{4 \pi} ~
		 p' ( \vartheta ', \varphi ')~ \hat{\vec{K'}} ~ d \Omega ' ~.
\end{equation}

The momentum (per unit time)
obtained by the particle due to the interaction
with radiation is
\begin{equation}\label{5}
\frac{d~ \vec{p'}}{d~ t} = \frac{1}{c} ~ S' ~ \left \{
			      C'_{ext} ~-~ C'_{sca} ~ \int_{4 \pi} ~
			      p' ( \vartheta ', \varphi ')~ \hat{\vec{K'}} ~
			      d \Omega ' \right \} ~.
\end{equation}
As for the energy, we suppose that $d E' / d t =$ 0.

For the sake of brevity, we will use ``effective factors'' $Q'_{xxx}$
instead of effective cross-sections $C'_{xxx}$:
$C'_{xxx} = Q'_{xxx} ~ A'$, where $A'$ is cross-section of a sphere of
volume equal to the volume of the particle.
Equation (5) can be rewritten to the form
\begin{eqnarray}\label{6}
\frac{d ~\vec{p'}}{d~ t} &=& \frac{1}{c} ~ S'~A'~ ~ \left \{ \left [
	     Q'_{ext} ~-~ < \cos \vartheta'> ~ Q'_{sca} \right ] ~
	     \hat{\vec{S}_{i}} ' ~+~ \right .
\nonumber \\
& &  \left . \left [ ~-~ < \sin \vartheta' ~ \cos \varphi ' > ~ Q'_{sca}
	     \right ] ~ \hat{\vec{e}_{1}} ' ~+~
	     \left [ ~-~ < \sin \vartheta' ~ \sin \varphi ' > ~ Q'_{sca}
	     \right ] ~ \hat{\vec{e}_{2}} ' \right \} ~,
\end{eqnarray}
or, in a short form
\begin{equation}\label{7}
\frac{d~ \vec{p'}}{d~ t} = \frac{1}{c} ~ S'~A'~ ~ \left \{
	     Q_{R} ' ~ \hat{\vec{S}_{i}} ' ~+~ Q_{1} '
	     ~ \hat{\vec{e}_{1}} ' ~+~ Q_{2} '
	     ~ \hat{\vec{e}_{2}} ' \right \} ~.
\end{equation}

\section{Stationary frame of reference}
By the term ``stationary frame of reference'' we mean a frame of reference
in which particle moves with a velocity vector $\vec{v} = \vec{v} (t)$.
The physical quantities measured in the stationary frame of reference
will be denoted by unprimed symbols.

Our aim is to derive equation of motion for the particle in the
stationary frame of reference.
We will use the fact that we know
this equation in the proper frame of reference -- see Eq. (7) -- and
$d E' / d t =$ 0, which corresponds to the fact that particle's mass $m$
does not change -- $d m / d t =$ 0.

\subsection{Flux density of the radiation energy}
As for the relation between $S'$ and $S$, we refer the reader to
Eqs. (69), (113) or (116) in Kla\v{c}ka (1992), or, to
Eqs. (9) in Kla\v{c}ka (1993a). We may write
\begin{equation}\label{8}
S ' = S ~ ( 1 ~-~ 2 ~ \vec{v} \cdot \hat{\vec{S}_{i}}  / c ) ~.
\end{equation}

\subsection{Aberration of light}
As for the transformation between the unit vectors $\hat{\vec{S}_{i}} '$
and $\hat{\vec{S}_{i}}$, it is given by the well-known phenomenon
called aberration of light. The result is (see, e. g.,
Eq. (19) in Kla\v{c}ka (1993a)):
\begin{eqnarray}\label{9}
\hat{\vec{S}_{i}} ' &=&  ( 1 ~+~ \vec{v} \cdot \hat{\vec{S}_{i}}  / c ) ~
		       \hat{\vec{S}_{i}} ~-~ \vec{v} / c ~,
\nonumber \\
\hat{\vec{e}_{j}} ' &=&  ( 1 ~+~ \vec{v} \cdot \hat{\vec{e}_{j}}  / c ) ~
		       \hat{\vec{e}_{j}} ~-~ \vec{v} / c ~, ~~ j = 1, 2 ~.
\end{eqnarray}
The inverse transformation yields
\begin{eqnarray}\label{10}
\hat{\vec{S}_{i}}  &=&	( 1 ~-~ \vec{v} \cdot \hat{\vec{S}_{i}} ' / c ) ~
		       \hat{\vec{S}_{i}} ' ~+~ \vec{v} / c ~.
\nonumber \\
\hat{\vec{e}_{j}}  &=&	( 1 ~-~ \vec{v} \cdot \hat{\vec{e}_{j}} ' / c ) ~
		       \hat{\vec{e}_{j}} ' ~+~ \vec{v} / c ~, ~~ j = 1, 2 ~.
\end{eqnarray}

\subsection{Transformation of force}
We can immediately write
\begin{equation}\label{11}
\frac{d ~ \vec{p}}{d ~t} = \frac{d ~ \vec{p} '}{d ~t} ~.
\end{equation}

\subsection{Equation of motion}
Putting Eqs. (7) , (8) and (9) into Eq. (11), we finally obtain
\begin{eqnarray}\label{12}
\frac{d~ \vec{v}}{d ~t} &=&  \frac{S ~A'}{m~c} ~ \left \{ Q_{R} ' ~ \left [
		 \left ( 1~-~ \vec{v} \cdot \hat{\vec{S}_{i}} / c \right ) ~
		 \hat{\vec{S}_{i}} ~-~ \vec{v} / c \right ] ~+~ \right .
\nonumber \\
& &  \left .  \sum_{j=1}^{2} ~Q_{j} ' ~\left [  \left ( 1~-~ 2~
	      \vec{v} \cdot \hat{\vec{S}_{i}} / c ~+~
	      \vec{v} \cdot \hat{\vec{e}_{j}} / c \right ) ~ \hat{\vec{e}_{j}}
	      ~-~ \vec{v} / c \right ] \right \} ~.
\end{eqnarray}

\section{Conclusion}
We have derived equation of motion for real dust particle under the action
of electromagnetic radiation. Equation of motion is
represented by Eq. (8) in the proper frame of reference of the
particle, and, by Eq. (12) in the stationary inertial frame of reference
in which particle's immediate velocity is $\vec{v}$. All the results
may be applied to real motions if the accuracy
to the first order in $\vec{v} / c$ is sufficient.
As for practical applications, the terms $v/c$ standing at $Q_{j}'$
($j =$ 1, 2)
are negligible for majority of real particles.
(We want to stress that values
of $Q'-$coefficients depend on particle's orientation with respect to the
incident radiation -- their values are time dependent.)

\acknowledgements
Special thanks to the firm ``Pr\'{\i}strojov\'{a} technika, spol. s r. o.''.
The paper was partially
supported by the Scientific Grant Agency VEGA (grant No. 1/7067/20).

\end{document}